\documentclass[useAMS,usenatbib,usegraphicx]{mn2e}
\usepackage{amsmath}

\title[Solar-like rotation periods with Gaia]
 {Determination of rotation periods in solar-like stars with irregular sampling: the Gaia case}
\author[E. Distefano et al.]
{\parbox{\textwidth}{E.~Distefano,$^{1}$\thanks{E-mail: \texttt{Elisa.Distefano@oact.inaf.it}}
A.~C.~Lanzafame,$^{2,1}$ 
A.~F.~Lanza,$^1$ 
S.~Messina,$^1$ 
A.~J.~Korn,$^3$ 
K.~Eriksson,$^3$ 
J.~Cuypers$^4$}\vspace{0.4cm} \\
\parbox{\textwidth}{$^1$INAF - Catania Astrophysical Observatory, via S. Sofia 78, 95123 Catania, Italy\\
$^2$University of Catania, Astrophysics Section, Dept. of Physics and Astronomy, via S. Sofia 78, 95123 Catania, Italy\\
$^3$Division of Astronomy and Space Physics, Department of Physics and Astronomy, Uppsala University, Box 516, 75120, Uppsala, Sweden\\
$^4$Royal Observatory of Belgium, Ringlaan 3, 1180 Brussel, Belgium
  }}
\date{Released 2011 Xxxxx XX}

\pagerange{\pageref{firstpage}--\pageref{lastpage}} \pubyear{2011}

\def\LaTeX{L\kern-.36em\raise.3ex\hbox{a}\kern-.15em
    T\kern-.1667em\lower.7ex\hbox{E}\kern-.125emX}

\begin{document}

\label{firstpage}

\maketitle

\begin{abstract}
We present a study on the determination of rotation periods ($P$) of solar-like stars from the photometric irregular time-sampling of the ESA Gaia mission, currently scheduled for launch in 2013, taking into account its dependence on ecliptic coordinates. 
We examine the case of solar-twins as well as thousands of synthetic time-series of solar-like stars rotating faster than the Sun.
In the case of solar twins we assume that the Gaia unfiltered photometric passband $G$ will mimic  the variability of the total solar irradiance (TSI) as measured by the VIRGO experiment.
For stars rotating faster than the Sun, light-curves are simulated using synthetic spectra for the quiet atmosphere, the spots, and the faculae combined by applying semi-empirical relationships relating the level of photospheric magnetic activity to the stellar rotation and the Gaia instrumental response.
The capabilities of the Deeming, Lomb-Scargle, and Phase Dispersion Minimisation methods in recovering the correct rotation periods are tested and compared. 
The false alarm probability (FAP) is computed using Monte Carlo simulations and compared with analytical formulae.
The Gaia scanning law makes the rate of correct detection of rotation periods strongly dependent on the ecliptic latitude ($\beta$). 
We find that for $P \simeq$ 1\,d, the rate of correct detection increases with $\beta$ from 20-30 per cent at $\beta \simeq 0$ to a peak of 70 per cent at $\beta=45^{\circ}$, then it abruptly falls below 10 per cent at $\beta > 45^{\circ}$.
For $P > 5$\,d, the rate of correct detection is quite low and for solar twins is only 5 per cent on average.

\end{abstract}

\begin{keywords}
 stars: rotation -- stars: late-type -- methods: data analysis -- surveys -- open clusters and associations: general.
\end{keywords}

\section{Introduction}

Main-sequence stars with spectral types later than F5 show variability phenomena due to solar-like magnetic activity. As in the Sun, magnetic fields generated in the convection zone produce active regions (ARs) consisting of cool spots and bright faculae. 
The visibility of the ARs is modulated by the rotation of the star and the associated modulation of the optical flux induced by their brightness inhomogeneities can be used to derive its rotation period. 
Nevertheless, active regions are far from being ideal tracers for stellar rotation studies because they are subject to an intrinsic evolution on different timescales, ranging in the Sun from a few hours to several months in the case of the activity complexes. 
More precisely, \citet{lanza04} studied the typical time-scales of the solar variations by  analysing time-series of the total solar irradiance (TSI) and the spectral irradiance at 402, 500, and 862 nm (SSIs) as observed by the VIRGO experiment on board of the SoHO satellite from 1996 to 2002.
They showed that the intrinsic evolution of sunspots occurs on a typical time-scale $\tau_s \simeq 9$ d whereas the evolution of the faculae has a time-scale of the order $\tau_f \simeq 60$ d. 
The variability due to the rotational modulation has a time-scale set by the solar rotation period and is of the order of 30~d.
On time intervals up to 60~d the variation of the solar flux is therefore dominated by the intrinsic evolution of ARs and the rotational modulation.
Variations on a longer time-scale, up to 200 d, are attributed to the intrinsic
evolution of ARs complexes at "active longitudes" i.e. heliographic longitudes characterised by the frequent, localised emergence of new magnetic flux \citep[e.g.,][]{dona}.
Finally, at even longer time-scales, the flux variations are mainly due to the 11-yr sunspot cycle.

The study of variability phenomena in solar-like stars of different ages and spectral types is crucial to understand how the stellar rotation and the magnetic activity evolve in time and how they are related to each other and to the global stellar parameters. 
The analysis of long-term photometric time-series of solar-like stars in the optical passbands, like those that will be provided as part of Gaia mission data-flow, is a powerful tool to investigate the variability induced by magnetic activity.
The aim is, in the first place, to extract rotation periods from the rotational modulation and to correlate them with the magnetic activity level inferred by the light-curve amplitude (e.g., \citealt{lanza01};  \citealt{lanza06b}; \citealt{messina02};  \citealt{messina08}). 
The rotational modulation can be used to further constrain the geometrical and physical properties of ARs (e.g., \citealt{lanza01}; \citealt{messina06}). 
Observations during different seasons can further reveal information on the surface differential rotation, since the ARs latitudes change in time because of the magnetic cycle. 
Finally, long-term flux variation can allow us to estimate magnetic cycles' time-scales \citep[e.g.,][]{messina02}.

Gaia is an ESA (European Space Agency) mission that  will perform a multiepoch survey of the whole sky  with the aim to chart a three-dimensional map of the Milky Way  (\citealt{perry}; \citealt{mig}). 
The mission, scheduled for launch in 2013, will have a duration of about 5 years and will supply astrometric, photometric, and spectro-photometric  measurements for about 1 billion sources down to the limit magnitude $V~\approx~20$. 
A crude estimate suggests that about 20 millions of the observed sources will be detected as variables \citep{EC00}. 
Therefore this survey offers  a great opportunity to  broaden significantly the statistics and the characterisation of solar-like variables. 
At the end of the Gaia mission a photometric time-series in a broad unfiltered passband (the so called  G-band) with a time baseline of about 1800 d will be available for each source observed by the satellite.
The number of per-source observations and, consequently, the number of points for each time-series, depends on  the source coordinates and ranges from about 40 to 250  with a mean value of 80.

\citet{EMI} studied the rate of correct detection of periods with Gaia as a function of the ecliptic coordinates for strictly periodic variable stars.
They performed an extensive analysis by simulating thousands of  light curves characterised by different periods and  signal-to-noise ratios.
They showed that the rate of detection depends mainly on the ecliptic latitude of the simulated sources and that, for a signal-to-noise ratio $S/ N~\ge~1.5$, it is close to 100 per cent for all simulated periods.

The variability simulated by \citet{EMI} consists, however, of a simple sinusoidal signal with a fixed frequency so their results do not apply to solar-like variability. Indeed, light-curves of solar-like stars are more complex than a simply sinusoidal signal because the amplitude and phase of the flux modulation change in time owing to the intrinsic evolution of ARs and AR complexes (see e.g. \citealt{messina02}, \citealt{messina03}). 
In the present work, we simulate light-curves of solar-like stars characterised by different rotation periods and magnitudes, and analyse  Gaia's capability of correctly detecting rotation periods.
Moreover, we tested the efficiency of different period search algorithms in order  to find out the method that gives the best performance  for this kind of variability given the peculiar Gaia sampling.  
In Sect. 2   we give a brief description of the Gaia  instruments and illustrate the main features of the Gaia  scanning law.
In Sect. 3 we describe the methods  used to simulate Gaia photometric time-series for solar twins and for solar-like stars with a rotation period ranging from 0.3 d to 180 d. 
In Sect. 4 we discuss the results obtained by running the period search algorithms on the simulated time-series.

\section{The Gaia mission}

\subsection{The instruments}

Gaia will perform its observations from a controlled Lissajous-type orbit around the L2 Lagrange point of the Sun and Earth-Moon system.
Gaia has two telescopes with two associated viewing directions with a field of view of $0.7^{\circ} \times 0.7^{\circ}$ each.
The two viewing angles are separated by a highly-stable Ôbasic angleÕ of $106.5^{\circ}$.
The two field of views are combined into a single focal plane covered with CCD detectors. By measuring the instantaneous image centroids from the data sent to ground, Gaia measures the relative separations of the thousands of stars simultaneously present in the combined fields. 

The astrometric field (AF) in the focal plane is sampled by an array of 62 CCDs, each read out in TDI (time-delayed integration) mode, synchronised to the scanning motion of the satellite. 
Stars entering the combined field of view first pass across dedicated CCDs which act as a Õsky mapperÕ (SM) - each object is detected on board and information on its position and brightness is processed in real-time to define the windowed region read out by the following CCDs.
Before stars leave the field of view, spectra are measured in three further sets of dedicated CCDs. The BP and RP CCDs - BP for Blue Photometer and RP for Red Photometer - record low-resolution prism spectra covering the wavelength intervals 330-680 and 640-1000 nm, respectively.
In addition to the low-resolution spectro-photometric instrument, Gaia features a medium-resolution integral-field spectrograph, the so-called Radial Velocity Spectrometer (RVS) instrument.

The astrometric measurements will be performed in the  G-band, i.e. a broad unfiltered passband that covers the wavelength range from about 350 to 1000 nm, with the  maximum  transmission at $\sim$ 715~nm and a full width at half maximum of 408 nm \citep{jordi}.
The expected photometric precision for a single measurement is better than 2~mmag for  targets brighter than $G$ = 15 and about 10~mmag for the faintest sources (G~$\sim$20).  
All data  will be analysed through  software packages written in the framework of the Gaia DPAC (Data Processing and Analysis Consortium).

\subsection{The Gaia scanning law}

\begin{figure}
\includegraphics[angle=-90,width=85mm]{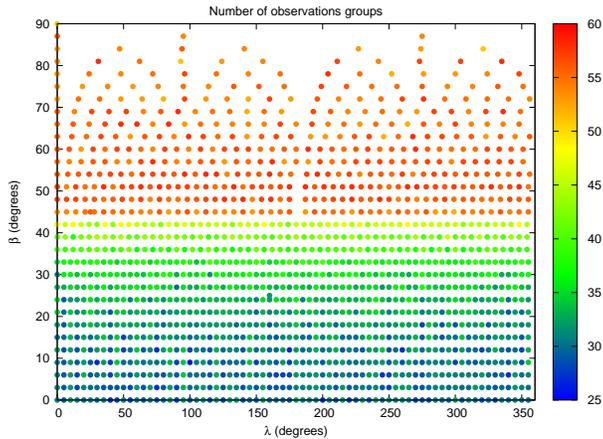}
\caption{Number of observation groups (short-term sequences of observations closely spaced in time) at different ecliptic coordinates. 1500 coordinate pairs are considered. 
Two different regimes are clearly visible: the average number of groups is about 50 at $\beta \ge 50^{\circ}$ and about 30 for $\beta \le 40^{\circ}$.}
\label{ngroups}
\end{figure}

During its 5-year operational lifetime, the Gaia satellite will continuously spin around its axis at a constant speed of 60~arcsec~s$^{-1}$.
Therefore, over a period of 6 hours, the two astrometric fields of view will scan across all objects located along the great circle perpendicular to the spin axis.
Because of the basic angle of 106.5$^{\circ}$ separating the astrometric fields of view on the sky, objects transit the second field of view with a delay of 106.5 minutes compared to the first field.
The spin axis, maintained at a fixed angle $\xi = 45^{\circ}$  with respect to the solar direction, is in turn characterised  by a slow precession  (about $4^{\circ}$ per day) around the Sun, so that the great circle mapped by the telescopes slowly changes in time.
As a result of the scanning law, the time sampling in the G-band is quite irregular and the number of observations (transits), for a given source,  is a function of the ecliptic latitude.
Basically, each time-series is made  by several groups of observations, i.e. short-term sequences of observations closely spaced in time. 
\citet{EMI} studied  the features of the Gaia scanning law in details and  showed that two different regimes can be identified:
time-series  will consist of about 30 groups of observations for sources placed at ecliptic latitudes $-30^{\circ} \le \beta \le 30^{\circ}$,  whereas in the directions  above $\beta = 40^{\circ}$ and below $\beta= -40^{\circ}$ the average number of groups of transits is 50.
Such regimes are clearly visible in Fig. \ref{ngroups} where we plotted the number of observation groups $N_G$ for 1500 sky directions uniformly distributed in the hemisphere delimited  by $0^{\circ} \le \lambda \le 360^{\circ}, 0^{\circ} \le \beta \le 90^{\circ}$.

\begin{figure}
\includegraphics[angle=-90,width=85mm]{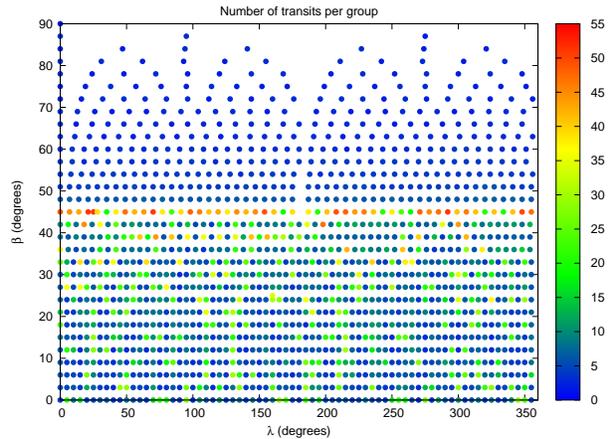}
\caption{Maximum number of observations (transits) per group for the same coordinate pairs of Fig.\,\ref{ngroups}.} 
\label{npoints}
\end{figure}

The number of observations per group has a more complicated  dependence on the ecliptic coordinates. At latitudes $\beta~\le~-45^{\circ}$ and $\beta~\ge~45^{\circ}$ a group consists, on average, of only two observations spaced by a six-hours interval corresponding to two consecutive satellite revolutions.
At latitudes $-45^{\circ}~\le~\beta \leq 45^{\circ} $, the number of observations per group ranges between 2 and 60. 
In Fig. \ref{npoints} we plot the maximum number of transits per group for each of the ecliptic coordinate pairs considered.

\section{Simulations}

In this section we illustrate the methods used to simulate photometric time-series in the Gaia G-band for solar twins and for solar-like stars rotating faster than the Sun with different apparent magnitudes.
All simulated sources are placed at different ecliptic coordinates  to study the rate of correct period detection as a function of the observing direction.

\subsection{Simulation of photometric time-series of solar-twins}

The total solar irradiance variability closely mimics the changes of the solar flux in the optical band (e.g., \citealt{fro97}, \citealt{fli98}).
Therefore, to first order, we can assume that the TSI  variations, as measured by the VIRGO experiment on board of the SoHO satellite, can be taken as a proxy for the flux variations as measured by Gaia in the integrated G-band for a solar-twin star. 
The TSI time-series used in the present work consists of one measurement per hour, each measurement having a relative accuracy of $2.0 \times~10^{-5}$.
This time-series is the level 2.0 series of TSI made available by the VIRGO team on the official web site\footnote{http://www.pmodwrc.ch/. The measurements are listed in the file virgo\_tsi\_h\_v6\_001\_0803.dat} of the instrument. 
The TSI measurements we considered span the time interval between 7 February 1996 and 5 March 2008 and, therefore, overlap with a complete solar activity-cycle.
They show occasional gaps with typical durations of a few days, except for the four-months gap in mid 1998, when the spacecraft tracking was lost.
Details on the procedure applied to obtain level 2.0 data can be found in \citet{ank99}, \citet{fro01} and \citet{fro03}.

\begin{figure}
\includegraphics[angle=-90,width=85mm]{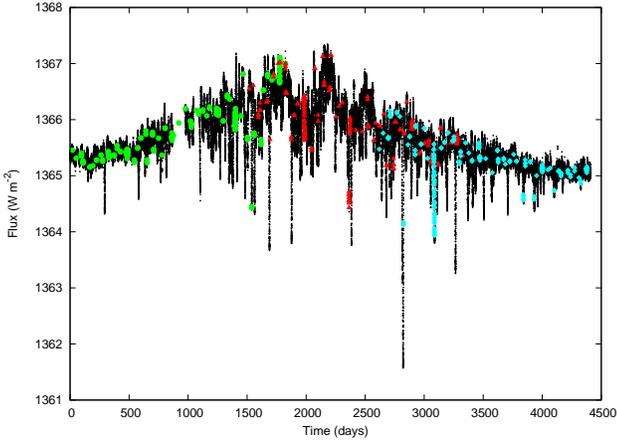}
\caption{The TSI time-series as measured by the VIRGO experiment (black dots). A 5-yr (mission duration) window sampled with the Gaia scanning law at $(\lambda =45^{\circ}, \beta=45^{\circ})$ is over-plotted on the TSI measurements in order to simulate photometric time-series acquired by Gaia in the G-band. The window is shifted along the TSI time-series to cover different phases of the solar cycle (green bullets, red triangles and blue diamonds).}
\label{tsigaia}
\end{figure}

For our purposes, we resampled the  TSI time-series  according to the Gaia scanning law at different values of ecliptic coordinates $(\lambda, \beta)$ to simulate Gaia time-series. 
We chose 770 directions uniformly distributed in the sky region delimited by $0^{\circ} \le \lambda \le 180^{\circ}$ and  $0^{\circ} \le \beta \le 90^{\circ}$.
We restricted our simulations to ecliptic latitudes $0^{\circ} \le \beta \le 90^{\circ}$ because the features of the Gaia sampling are quite symmetric with respect to the ecliptic plane.
The restriction in longitude is justified by the fact that the rate of correct period detections for strictly periodic variables  depends mainly on the ecliptic latitude and show only  little dependence on longitude \citep{EMI}.
For each chosen direction ($\lambda_0, \beta_0$), we  select a time interval of 5 yrs (the duration of the Gaia mission) in the TSI time-series and then choose, in such an interval, TSI data points according to the Gaia scanning law sampling at ($\lambda_0, \beta_0$).
The 5-yr window was then shifted with steps of 300 days along the entire TSI time-series.
In this way, for each chosen sky direction, we obtained 10 time-series  which sample different parts of the solar activity cycle with the Gaia time-sampling.
In Fig. \ref{tsigaia} three of the time-series generated at ($ \lambda =45^{\circ}, \beta = 45^{\circ}$) are over-plotted on the whole TSI time-series. 
The simulated time-series are characterised by the same number of points and time sampling, but cover different intervals of the solar variability along the 11-yr cycle.
 
The sampling at a given sky position has been generated through AGISLab (Astrometry Global Iterative Solution Laboratory), a package developed by  the  Coordination Unit 3 of the Gaia DPAC \citep{ho10}.
Finally, we degraded the time-series accuracy by adding the instrument noise (according to the error model described in \citealt{jordi})  to simulate  stellar targets with an average magnitude G=10. 
      
\subsection{Simulation of photometric time-series for solar-like stars with different rotation periods}
\label{sec:solar-like-simulations}

\subsubsection{The model}
We used the approach described by \cite{lanza06a} to generate synthetic light curves of solar-like stars. 
It assumes that the stellar variability can be modelled by means of three discrete active regions plus a uniform background component. 
This  is based on the work of \cite{lanza03} where the TSI time-series is fitted with this simple model with a residual standard deviations of 30-35  parts-per-million \citep[see also][]{bonomo08}.
Following \cite{lanza06a}, we simulate  the flux variations  at a given wavelength $\lambda$ and time $t$ as:

\begin{equation}
\label{model}
\begin{split}
\Delta & F(\lambda, t) \equiv F(\lambda,t) - F_u(\lambda) = \\  
 =& \sum_{k:\mu_k > 0 } \mu_k A_k I_u(\lambda,\mu_k)\left [c_{\rm spot}(\lambda) + Q c_{\rm fac}(\lambda, \mu_k) \right ]
 \\ 
 +& \Delta F_{\rm back}(\lambda, t),
\end{split}
\end{equation}

$F_u(\lambda)$ is the unperturbed flux (i.e. the monochromatic flux coming from the star in the absence of magnetic activity) and $F(\lambda, t)$ the perturbed flux at wavelength $\lambda$ and time $t$; 
$A_k$ is the area, at time $t$, of the cool spots associated with the $k$-th AR ($k~$=1,2 or 3).
The contribution of surface brightness inhomogeneities to the stellar flux is proportional to their projected area over the  stellar disk, so $A_k$ is multiplied by the factor $\mu_k \equiv \cos \psi_k$, where $\psi_k$ is the angle between the normal to the $k$-th AR, assumed to be point-like, and the line of sight;
$I_u(\lambda, \mu)$ is the specific intensity of the unperturbed photosphere, $c_{\rm spot}$ the contrast of cool spots, $c_{\rm fac}$ the facular contrast, $Q$ the ratio of the facular area to the spotted area in each active region (assumed to be constant) and $\Delta F_{\rm back}(\lambda, t)$ the flux variation due to the uniformly distributed background.
The contrast factors $c_{\rm spot}$ and $c_{\rm fac}$ depend on $\Delta T_{\rm spot}$ and $\Delta T_{\rm fac}$ respectively, where $\Delta T_{\rm spot}$ is the average temperature difference  between the unperturbed photosphere  and the spotted areas while $\Delta T_{\rm fac}$ is the temperature difference between faculae  and the unperturbed photosphere  (for details on the evaluation of $c_{\rm spot}$ and $c_{\rm fac}$, see \citealt{lanza06a}).  

\subsubsection{Simulation of synthetic spectral time-series for  solar-like stars}

Time-series of synthetic spectra for solar-like stars were simulated according to equation (\ref{model}).
The method, described in details in \citet{lanza06a}, provides spectra $F(\lambda,t)$ of a solar-like inhomogeneous photosphere by combining specific intensities $I(\lambda,\mu)$ computed assuming plane-parallel model atmospheres with effective temperature ($T_{\rm eff}$) equal to that of the unperturbed photosphere as well as with $T_{\rm eff}$ differing from the unperturbed photosphere by $\Delta T_{\rm spot}$ and $\Delta T_{\rm fac}$ in the spots and faculae, respectively.
The specific intensities used in this work were computed using of MARCS \citep{Gustafsson_etal:08} for $\log g=4.5$ and solar chemical composition. 
The parameters (areas and coordinates) of the three active regions and the background component $\Delta F_{\rm back}(\lambda, t)$ at a given time $t$ are computed by interpolating in time the values supplied from the best-fits of the TSI time-series \citep{lanza03}.
The dependence of the  rotational modulation amplitude on the rotation period $P$ and $T_{\rm eff}$ is taken into account by scaling the spot areas and the background component as described in \citet{lanza06a}.

\begin{figure}
\includegraphics[angle=-90,width=85mm]{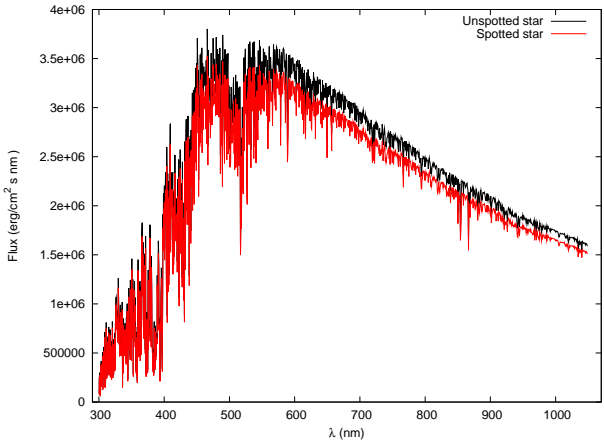} 
\caption{Comparison between spectra of an unperturbed (black) and a magnetically active (red line) photospheres.}
\label{spectrasimul}
\end{figure}

\begin{figure}
\includegraphics[angle=-90,width=85mm]{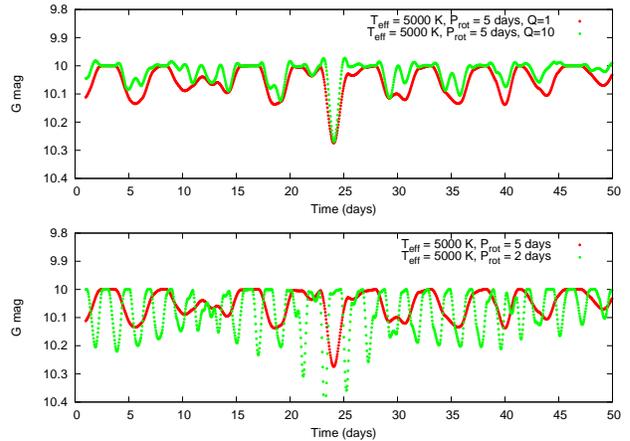}
\caption{Top panel: simulated light curves for two stars having the same effective temperature $T_{\rm eff}$ and the same rotation period but a different ratio $Q$ between facular and spotted areas. 
Bottom panel: two synthetic light curves for stars with the same effective temperature $T_{\rm eff}$ and  the same $\Delta T_{\rm spot}$ and $Q$ values  are plotted.} 
\label{outputsimul}

\end{figure}

As illustrative examples, in Fig. \ref{spectrasimul} we compare the simulated spectrum of a magnetically active star with the corresponding unperturbed spectrum and in Fig. \ref{outputsimul} we show the effects on the light-curve of varying $Q$ and $P$. Note that in stars whose variability is "faculae dominated", the rotational modulation is less evident because faculae produce an increase of the optical stellar flux that counteracts the effect of cool spots (Fig. \ref{outputsimul}, top panel). In stars rotating faster the  variability amplitude is larger (Fig. \ref{outputsimul}, bottom panel) because the magnetic activity is higher.

\subsubsection{Simulation of G-band time-series}

The procedure we follow to generate a G-band time-series for a star placed at ecliptic coordinates ($\lambda_0, \beta_0$) consists of the following steps:
\begin{itemize}
\item{we run the AGISLab code and compute the transit times at ($\lambda_0, \beta_0$) according to the Gaia scanning law;}
\item { for each transit time we compute the  spectrum  $\ F(\lambda, t)$ as given by equation\,(\ref{model});}
\item {we convolve the spectrum $ F(\lambda, t)$ with the transmittance profile of the Gaia $G$ band to obtain the stellar flux in the $G$-band;}
\item{we add noise to the light curve according to the error model described in \cite{jordi}.} 
\end{itemize}
We simulated photometric time-series in the G-band for about 1500 ecliptic coordinate pairs uniformly distributed in the region $0^{\circ}~\le~\lambda~\le~360^{\circ}$ and $ 0^{\circ}~\le~\beta~\le~90^{\circ}$.

The parameters used for our simulations are:
\begin{itemize}
\item Rotation period $P$: 0.3, 0.5, 1.0, 2.0, 4.0, 5.0, 7.5, 10.0, 20.0, 40.0, 60.0, 90, 180 $d$;
\item Effective temperature $T_{\rm eff}$= 5000 K;
\item Q =1;
\item $\Delta T_{\rm spot}$ =  -800 K;
\item $\Delta T_{\rm fac} $= 110 K;
\item Apparent magnitudes G= 10, 15, 19.
\end{itemize}

In total, we simulated about 58000 photometric time-series corresponding to $\sim~4.6~\times~10^6$ transits.

\section{Rate of correct period detections with Gaia}

In the present section we show the results obtained by running three period search algorithms on the simulated time-series.
The algorithms we tested are: the Deeming method \citep{dee75}, the Lomb-Scargle periodogram (\citealt{lo76}; \citealt{sca82}) and the Phase Dispersion Minimization method (PDM method; \citealt{ju71}; \citealt{ste78}).
We preliminary discuss  a method to estimate the FAP (False Alarm Probability) for the periods detected by the Lomb-Scargle algorithm which is the method  providing
the highest rate of correct detections.

\subsection{FAP estimation}

The FAP associated with a given peak with a power level  $z$ in the Lomb-Scargle periodogram is the probability that such a $z$ value is due to pure Gaussian noise,
which is the main error source for stars brighter than $G$ = 16 mag in the Gaia photometry.
In the general case Monte Carlo simulations can be used to estimate the FAP \citep[e.g.,][]{Frescura}.
These, however, are not suitable for the analysis of a very large dataset because of the computational load implied.
The Lomb-Scargle algorithm, however, allows an analytical estimate of the FAP which, according to \cite{fap}, is given by:
\begin{equation}
\label{FAP}
 \Pr[Z_{\max}~\ge~z]= 1 - \left[1 - \left(1 - \frac{z/\sigma_{\chi}^2}{N/2}\right)^{N/2}\right]^{M}
\end{equation}
for a given frequency $\nu$ with a power $z$ in the normalised Lomb-Scargle periodogram.
Equation\,(\ref{FAP}) gives the probability that the maximum peak power $Z_{max}$  exceeds a threshold  $z$  for a data set $\{\chi_i\}$ consisting of pure Gaussian noise;
$N$ is  the number of  data points used to compute the periodogram, $\sigma_{\chi}^2$  the variance of the data set $\{\chi_i\}$ and $M$  the number of  {\em independent frequencies}, i.e. the number  of frequencies  at which the periodogram powers are independent  variables.
If one normalises such that $\sigma^2_\chi = 1$, equation\,(\ref{FAP}) becomes

\begin{equation}
\label{FAPgaia}
\Pr[Z_{\max}~\ge~z]= 1 - \left(1 - \left(1 - 2z/N \right)^{N/2}\right)^{M}.
\end{equation}

Note that, as indicated by \cite{fap}, only in the limit $N \rightarrow \infty$, equation\,(\ref{FAP}) reduces to the \cite{hb86} expression: 
\begin{equation}
\label{FAPhorne}
 \Pr[Z_{\max}~\ge~z]= 1 - (1 -e^{-z})^{M} ,
\end{equation}
which is the distribution often adopted by others authors in this contest.

The difficulty  in applying  Eq.\,(\ref{FAP})  and Eq.\,(\ref{FAPhorne}) arises from  the unknown parameter $M$. 
In the case of a set of data with an even time sampling, the independent frequencies are given by 
\begin{equation}
 \nu_K~=\frac{k}{T},
\end{equation}
where $T$ is the time interval spanned by the data and $k=1,2,...N/2$. 
In this case $M=N/2$. 
In the case of a set of unevenly sampled data, $M$ cannot be derived analytically and can be inferred only by fitting empirical cumulative distribution functions (CDFs) of the peak powers $z$ generated by means of Monte Carlo simulations (see \citealt{hb86}, \citealt{Frescura}).  
One could expect that $M$ satisfies the condition $N/2\le M\le N_f$, where $N_f$ is the number of  inspected frequencies.
One possible choice is to adopt the value 
\begin{equation}
\label{eq:independent_frequency_prescription}
 M=\frac{N_f}{r}
\end{equation} 
where $r=(1/T)/\delta\nu$
and $\delta\nu$ is the frequency step used to sample the periodogram \citep{Frescura}. 
The factor $r$ is introduced to take into account that a periodogram is generally oversampled (i.e.  sampled with a frequency step $\delta\nu$ smaller than the minimum frequency interval $\Delta\nu_{\min}=1/T$ that can be resolved by the dataset). In our tests, we computed the periodogram in the frequencies domain (0.01 - 5 d$^{-1}$), with a frequency step $\delta\nu$ = 0.0001 d$^{-1}$ and, therefore, $N_f=49990$, $\Delta \nu_{\min} = 1/1800$ d$^{-1} = 0.00055$, $r=5$ and $M = N_f/r = 9980$. 
We used Monte Carlo simulations to construct an empirical FAP  for 8 different sky positions characterised by a different sampling and  number of observations and  made a comparison between such empirical distributions and  equation\,(\ref{FAPgaia}) computed adopting $M = 9980$.

\begin{figure*}
\includegraphics[angle=-90,width=170mm]{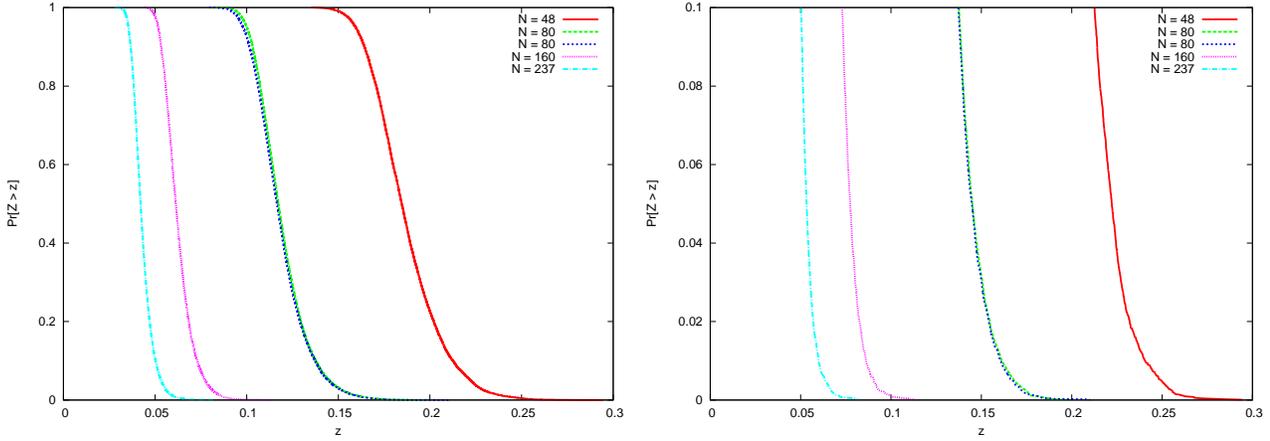}
\caption{Left panel: empirical CDFs constructed for five different sky positions ($\lambda,\beta$).
Right panel: enlargement of the critical region $\Pr[Z < z]<0.1$.  Note that the CDFs relative to two different sky positions, ($\lambda=0\degr, \beta= 75\degr$) and ($\lambda=160\degr, \beta=0\degr$), with the same number of observations but differently distributed appear to be very similar.}
\label{CDFs}
\end{figure*}

The approach we followed to generate the empirical FAP at a given position ($\lambda_0, \beta_0$) is the same described in \cite{Frescura}, i.e.:
\begin{itemize}
\item{we generated $10^4$ time-series of pure white  Gaussian noise  with the sampling given by the Gaia scanning law at ($\lambda_0, \beta_0$);}
\item{ we run the Lomb-Scargle algorithm on each  time-series by using the same grid of frequencies applied to period search in our tests; }
\item{for each of the $10^4$ periodograms we selected the  highest  power $Z_{max}$ and used these values to construct their cumulative distribution functions (CDFs).  }   
\end{itemize}
A CDF constructed in such a way can be regarded as an empirical representation of the probability function $\Pr[Z_{max}~\ge~z]$, i.e. the probability that pure noise alone could generate  a peak greater than or equal to a given threshold value $z$ in the Lomb-Scargle periodogram.
In Fig. \ref{CDFs} we show the CDFs constructed for five different sky positions.
The CDFs depend on the number of  observations, i.e. the lower  the number of  observations the higher  the probability of finding high $z$ peaks in a periodogram generated by a set of data consisting of  pure Gaussian noise.
In the same figure we also plotted the CDFs constructed for two different sky positions, ($\lambda=0\degr, \beta= 75\degr$) and ($\lambda=160\degr, \beta=0\degr$), in which the number of observations is the same 
($N = 80$) but the sampling is different, the first being more uniformly distributed, the second concentrated in 12 rather isolated groups. 
The two CDFs  are fairly similar, which indicates that the CDFs are much more sensitive to the number of observations than to the observations' distribution in time.

\begin{figure*}
\includegraphics[angle=-90,width=170mm]{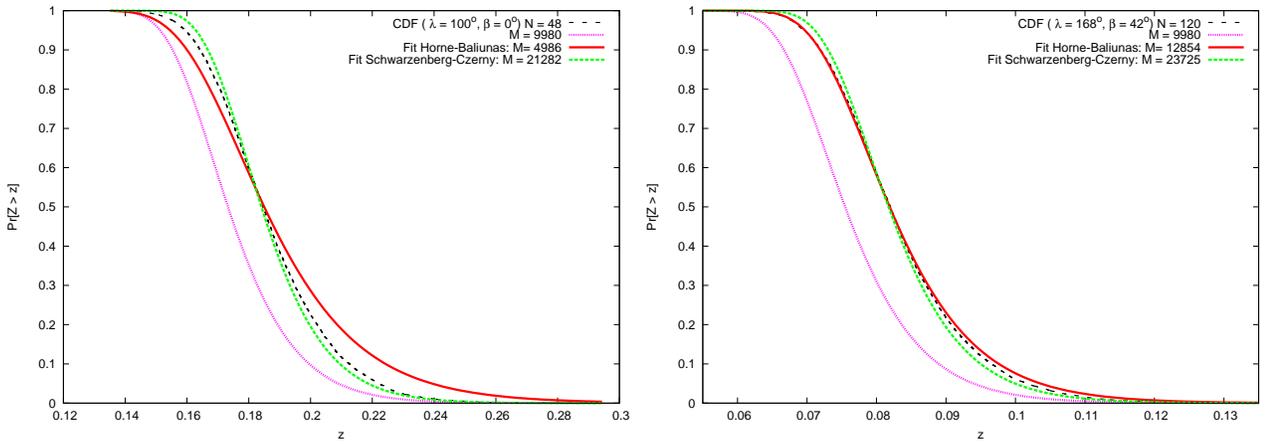}
\caption{A comparison between empirical CDFs (black dotted lines) and the analytical expression in equation (\ref{FAPgaia}) (in purple) computed with the number of independent frequencies evaluated using equation\,(\ref{eq:independent_frequency_prescription}).
Also plotted the results obtained by fitting the CDFs with equations\,(\ref{FAPgaia}) (in green) and (\ref{FAPhorne}) (in red) by varying $M$ as a free parameter.}
\label{Cuycdf}
\end{figure*}

In Fig.\,\ref{Cuycdf}, we compare the Monte Carlo CDFs with the probabilities given by equation\,(\ref{FAPgaia}) and the number of independent frequencies estimated by equation\,(\ref{eq:independent_frequency_prescription}), 
for two different sky directions.
The comparison shows that the FAP obtained using equations\,(\ref{FAPgaia}) and (\ref{eq:independent_frequency_prescription}) underestimates the true FAP.
In Fig.\,\ref{Cuycdf} we also plot the results obtained by fitting both equations (\ref{FAPgaia}) and (\ref{FAPhorne}) to the empirical CDFs using a non-linear least-squares (NLLS) Marquardt-Levenberg algorithm, varying $M$ as a free parameter. 
These best-fits indicate that the number of independent frequencies estimated using equation\,(\ref{eq:independent_frequency_prescription}) is too low and that a higher value should be adopted.
The shape of the CDFs is not accurately reproduced either by equation (\ref{FAPgaia}) or by equation (\ref{FAPhorne}), since the former tends to underestimate the FAP while the latter tends to overestimate it.

\begin{figure*}
\includegraphics[angle=-90,width=170mm]{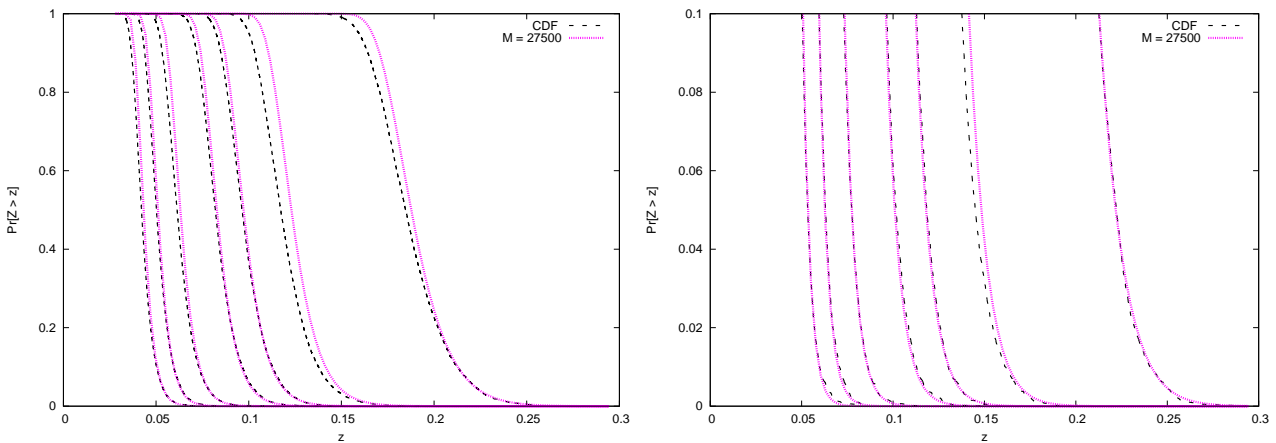}
\caption{Left panel: a comparison between empirical CDFs (black dotted lines) and the analytical expression (\ref{FAPgaia})  (in purple) computed by using $M = M_{\rm ad} = 27500$. 
Right panel: enlargement of the critical region $\Pr[Z < z] <0.1$.} 
\label{FAP27500}
\end{figure*}

In conclusion, an accurate FAP estimate would require the construction of CDFs for a convenient grid of sky coordinates. However as a first approximation, we shall use equation\,(\ref{FAPgaia}) adopting  a suitable number of independent frequencies $M$.
To do that, we fit equation\,(\ref{FAPgaia}) to the 8 empirical CDFs (corresponding to the selected 8 different sky positions) in the region $\Pr < 0.1$ (corresponding to a significance level $>90$ per cent), then we take the average of the resulting $M$ values, which gives $M_{\rm ad} = <M> =27500$, our adopted number of independent frequencies.
In Fig. \ref{FAP27500} we display the comparison between empirical CDFs and equation (\ref{FAPgaia}) computed by adopting $ M = M_{\rm ad}$. 
The empirical CDFs are not accurately reproduced in the region where FAP $>0.1$, but are reproduced with an accuracy better than $\sim$~5 per cent in the region where FAP~$<~0.1$ which is the region of interest for all practical purposes.

\subsection{Search for rotation period in solar-twins}

\begin{figure}
\includegraphics[angle=-90,width=85mm]{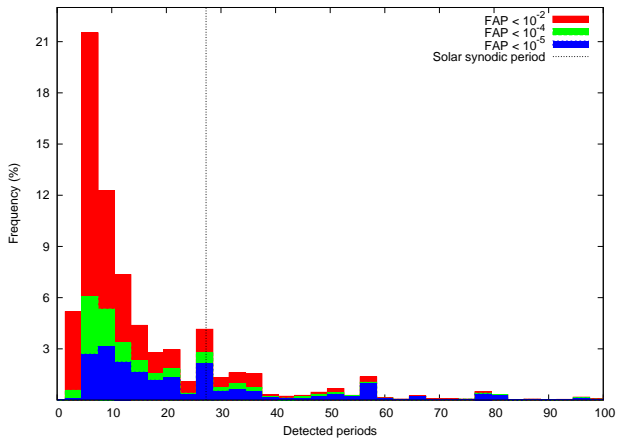}
\caption{Distributions of the periods as detected by Lomb-Scargle algorithm on simulated G-band time-series for a solar-twin star when time-series of 5-year extension are analysed.} 
\label{histotot}
\end{figure}

In Fig. \ref{histotot}, we show the distributions of periods  detected by using the Lomb-Scargle algorithm on solar-twin simulated time-series with a FAP $< 10^{-2}$, $10^{-4}$, and $10^{-5}$. 
For FAP $< 10^{-2}$, the true period (i.e. the solar synodic period $P_{\odot}~=27.27$~d) is recovered, with a relative error less than 10 per cent, only in a small fraction (about 5 per cent) of the 7700 simulated time-series, while in the 60 per cent of cases periods shorter than about 20 d are obtained. 
By decreasing the FAP threshold from $10^{-2}$ to $10^{-4}$, we obtain a minor decrease in the percentage of the recovered true period and a large decrease of the percentage of spurious periods below 20~d. 
These latter decrease below 15 per cent for FAP $<10^{-5}$, at the expenses of a diminished true period recovery at 3 per cent.

The results shown in Fig. \ref{histotot} are not  surprising. 
In solar-like stars, the  variability due to rotational modulation can be masked by the growth and decay of ARs and AR complexes.
Such variability phenomena are responsible for the other periods detected by the period search algorithms as shown in the distribution.
As we pointed out in the introduction, the time-scale of intrinsic evolution of a single AR in the Sun is comparable with the solar rotation period, but AR complexes at active longitudes have a lifetime of about $\tau_{\odot}~\simeq~200-250$~d \citep{lanza03}.
In other words, even though a single spot group has a short lifetime (one or two weeks),  AR complexes may persist for several solar rotations because new spot groups form as previous ones fade away.
Therefore, the solar flux variations due to the evolution of AR complexes mask the variations due to the rotation on  time-scales longer than $200-250$~d.
\citet{lanza04} showed that the rotation period of the Sun can be recovered only if long time-series are divided into segments with a time extension $\sim 150$~d to limit the effect of AR complexes evolution.
Moreover, the period was correctly detected only on segments falling close to the minimum  of the solar cycle.
In fact, during such a phase, the solar photosphere is dominated by faculae, which evolve on a  time-scale $\tau_{fac}~\simeq~60-80$~d~$\simeq~2-3~P_{\odot}$.
Therefore, close to the minimum of the 11-yr solar cycle, the signal induced by rotational modulation is more coherent than in the intermediate phase and the maximum, when the level of solar activity is higher and sunspots dominate the TSI variations. 
The sunspot evolution on a time-scale of 10-15 d, on the other hand, makes it difficult to detect the rotational modulation on a time-scale of 27 d.
Variability with a periodicity of the order of half the solar synodic rotational period may be due to the (still debated) presence of active longitudes \citep[see, e.g.][and references therein]{Balthasar:2007}, but the frequency of periods around this value in Fig.\,\ref{histotot}, even for FAP $<10^{-5}$, is too high to be consistent with the spectral power (relative to the synodic period) obtained by specific studies not limited by the Gaia sampling.
Therefore, the relatively high frequency of periods below 20~d are likely due to the combination of the lack of coherence of the signal and the peculiarities of the Gaia sampling.
Note also the presence of a persistent peak at periods around 58~d, which is not associated to any known variability time-scale.
Further detailed analysis on these issues is required but deferred to a future work.

The Deeming and PDM method give approximately the same rate of correct detections and are not discussed in details here.

\subsubsection{Time-series segmentation for solar-twins}

Keeping in mind the results of \citet{lanza04}, we tried  to increase the rate of correct period detection by splitting  each of the 7700 simulated time-series into sub-series (segments)  with a shorter extension $\tau$. 

\begin{figure}
\includegraphics[angle=-90,width=85mm]{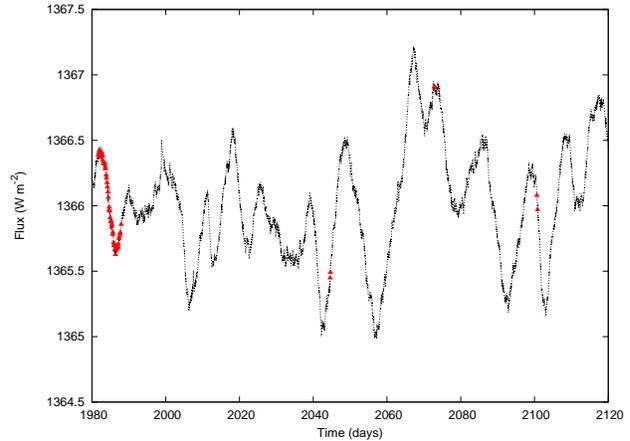}
\caption{A $160$~d sub-series extracted at ($\lambda = 45^\circ, \beta = 45^\circ$). In spite of the high number of observations, the flux modulation is poorly sampled because most of the points are distributed along an interval of only  $7~d$.}
\label{ingrandimento}
\end{figure}

The choice of $\tau$ is quite critical because it should be of the same order of the time-scale of AR complexes evolution but, at the same time, long enough to include a  sufficient number of data points. 
In our case, to detect the rotational modulation of a slowly rotating star like the Sun, the length of the sub-series should be selected in order to maximise the number of groups $N_G$ per sub-series rather than the number of single observations $N$.
In Fig. \ref{ingrandimento} we show a sub-series, extracted from one of the simulated time-series at ($\lambda = 45^\circ, \beta = 45^\circ$), with  a baseline $\tau~=160$~d (the same value used by \citealt{lanza04}) and consisting of a rather large number of single observations ($N=55$) but mostly concentrated on a single group spanning a 7-d interval. Clearly, such a kind of  sampling is inadequate to trace the solar flux modulation with a period  $P_{\odot}~=27.27$ d.

\begin{figure}
\includegraphics[angle=-90,width=85mm]{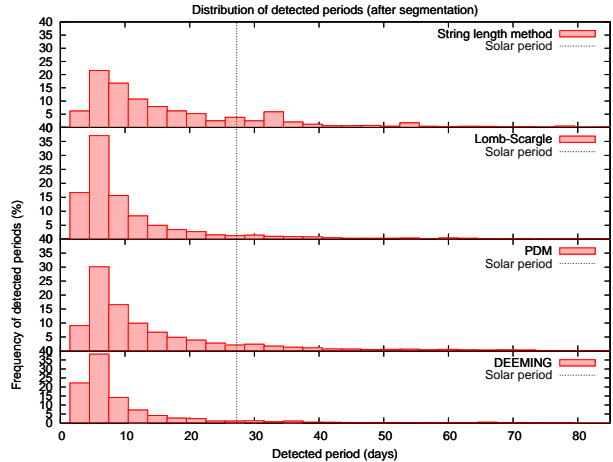}
\caption{Distribution of detected periods after applying time-series segmentation  }
\label{fre_450}
\end{figure}

The typical interval from two successive groups is about 45~d \citep{EMI}, thus we decided to use a baseline $\tau~=450$ d in order to generate sub-series including, on average, 10 groups of observations each.
We split each of the 7700 simulated time-series into 450 d segments  and run the period search algorithms  on all sub-series satisfying the condition $N_G \ge 10$.
We processed about 110000 sub-series and  plotted the distributions of detected periods in Fig. \ref{fre_450}.
Unfortunately, in our case, time-series segmentation has the effect to reduce the rate of correct detections.
We also tried to process the sub-series with the string-length method \citep{string}, that is particularly suitable for dealing with time-series with a low number  of points. 
Such a method is based on the minimisation of a quantity, called string-length, that is the sum of the lengths of the segments joining two successive points in the phase diagram.
This method is slightly more effective than the others but the rate of correct detections is always of the order of 5 per cent.

\begin{figure}
\begin{center}
\includegraphics[angle=-90,width=85mm]{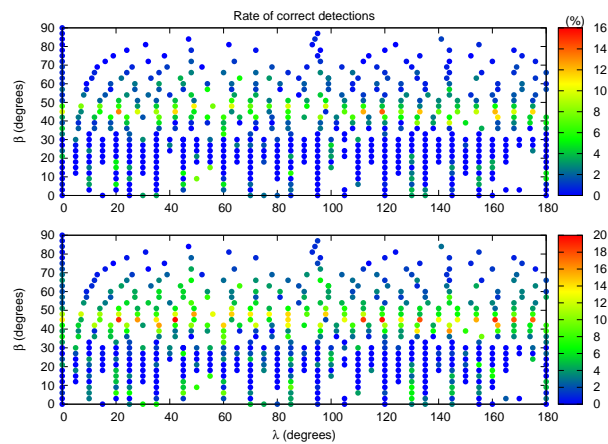}
\caption{Rate of correct detections with the string-length method: the top-panel refers to the case in which the solar period is detected as the first minimum of the periodogram whereas the bottom panel refers to the case in which the solar period coincides with one of the three periods with the highest probability.}
\label{mappa450}
\end{center}
\end{figure}

Finally, we analysed how the percentage of correct detections depends on the ecliptic coordinates of the simulated source. 
In Fig. \ref{mappa450}, we show the results obtained by running the string-length method on sub-series. 
For each of the simulated sky directions, the percentage of sub-series in which the rotation period of the Sun has been recovered with a relative error $\le~5$ per cent is plotted. 
The picture shows that the rate of correct detections (that has an average value $\sim~5$ per cent) is not uniformly distributed and peaks at ecliptic latitudes close to $\beta~=~45\degr$ where it reaches $\approx~15-20$  per cent. 
The detection rate obtained by running the other period search methods on sub-series have a trend similar to that shown in Fig. \ref{mappa450} and are not shown here. 

The main results of our analysis on solar twins can be summarised as follows:
\begin{itemize}
\item{the rate of correct period detection for solar-twins is generally very low and has an average value $\sim~5$ per cent; }
\item{segmentation of sub-series does not significantly increase the rate of correct detections;}
\item{at ecliptic latitudes close to $\beta~=~45\degr$ the rate of correct detections rises up to  15-20 per cent.}
\end{itemize}

\subsection{Search for rotation period in solar-like stars}

In this section, we investigate the Gaia's capability to detect rotational modulation  in solar-like stars with different rotation periods.
For doing that, we simulated photometric time-series in the Gaia G-band for solar-like stars placed at different ecliptic latitudes as described in Sect.\,\ref{sec:solar-like-simulations} with a rotation period ranging from 0.3 to 180~d and processed then with the Deeming method \citep{dee75}, the Lomb-Scargle periodogram (\citealt{lo76}; \citealt{sca82}) and the Phase Dispersion Minimization method (PDM method; \citealt{ju71}; \citealt{ste78}, with 5 bins in the phase diagram).

\begin{figure*}
\includegraphics[angle=-90,width=170mm]{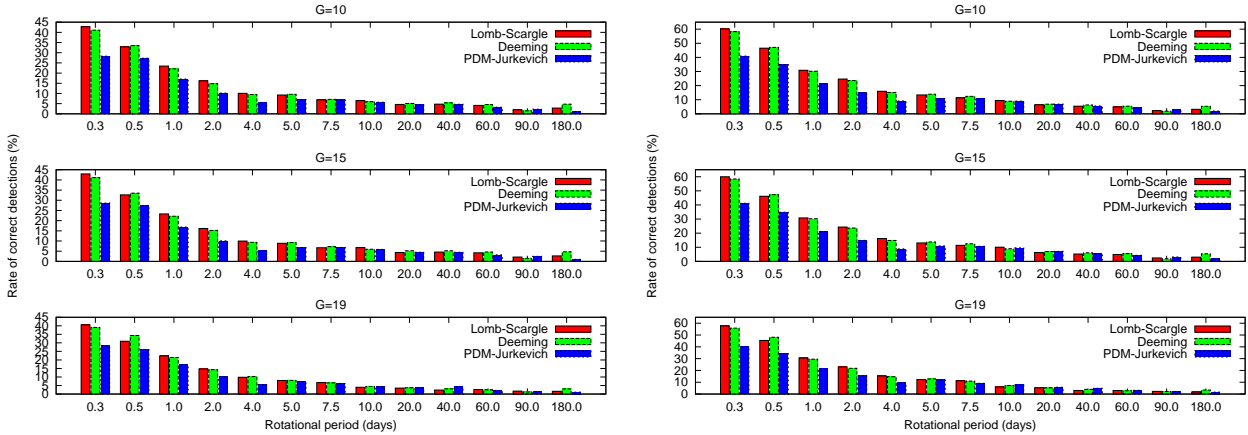}
\caption{Rate of correct period  detection on simulated light curves: the percentage of detections is a decreasing function of the rotation period; different colours refer to different period-search methods as labelled.
The rate of detections at G=10, 15 and 19 mag are shown in the top, medium and bottom panels respectively.
The left panels refer to the percentage of cases in which the period is correctly detected with a relative error $\le~5$ per cent whereas the histograms on the right panel indicate the percentage of cases in which rotation period is measured with a relative error  $\le~10$ per cent } 
\label{HISTO}
\end{figure*}

The histograms shown in Fig. \ref{HISTO} visualize, for each period search method, the percentage of cases in which the rotation period of the simulated stars  is correctly detected.
Histograms in the left panel refer to the cases in which the  rotation period is measured with a relative error $\le~5$ per cent  whereas those in the right panel refer to the cases in  which the rotation period is derived with a relative error $\le~10$ per cent. This percentage decreases further with decreasing relative errors limit below 5 per cent (not shown in figure), which indicates that, in our case, noise causes a rather wide spread of measured period around the true value. As discussed in Sect. 4.3.1, the active region evolution also contributes to the period spread around the true value (see Fig. \ref{confronto}).

The main results of our analysis are:
\begin{itemize}

\item{The rate of correct detections  decreases with increasing  rotation period and ranges from about 60 per cent for the fastest rotating stars ($P~=0.3$~d) to about 5 per cent for stars with a rotation period $P~\ge~20$~d;}

\item{The Deeming and  Lomb-Scargle methods are more effective than the PDM-Jurkevich algorithm.
The rotation period $P~=0.3$~d, for example, is recovered in about 60 per cent of the simulated sky directions by running either the Lomb-Scargle or the Deeming algorithm, whereas it is recovered only in $\approx~40$ per cent of the cases with the PDM-Jurkevich method;}

\item{For stars with a rotation period $P~\le~5$~d, the rate of correct detection does not depend on the apparent magnitude.
In fact  the amplitude of the light modulation is an increasing function of the angular velocity and, for $P~\le~5$~d, it ranges between $0.04-0.3$~mag  (e.g. \citealt{messina10a}; \citealt{messina10b}). 
Therefore the rotational modulation can be detected also at $G~\approx~19$ where the photometric error is $\sim~20$~mmag.}

\end{itemize}

\subsubsection{Time-series segmentation for fast-rotators}

Our knowledge on time-scales of AR growth and decay (ARGD)  in solar-like stars in general is still quite limited.
\cite{Donahue_etal_1997a}, from the analysis of 35 lower-main-sequence stars observed at Mount Wilson Observatory, estimated  the time-scale of active region evolution is approximately 50 days, while the lifetime of active region complexes is on the order of one year. 
Most of the stars in their sample have ARGD time-scales significantly longer than the rotation periods.
\cite{Donahue_etal_1997b} extended  such  analyses to 100 stars, showing that the variability of older stars (with lower activity and slower rotation) tend to be AR-evolution dominated, while younger stars (more active and fast rotating) have AR evolution time-scales significantly longer than the rotation period. 
The age at which this transition takes place also depends on the stellar colour (see Fig. 5 of \citealt{Donahue_etal_1997b}) and it is around the Sun's age or older for stars with $(B-V)$ similar to that of the Sun. Subsequently, \cite{messina03} analysed six young solar analogues and found that for four of them the time-scale at which the evolution of active regions begins to affect the variance of the observed flux ranges from $\simeq$5 months ($\simeq$30 rotation periods) to $\simeq$16 months ($\simeq$180 rotation periods).
Two of them were found AR evolution-dominated, in which case the active region evolution is not sufficiently distinct from the rotation period.

For the analysis of rotation periods in stars of young loose associations, \cite{messina10b,messina11} adopted a time-series segmentation of less than 60 days and used the comparison of the periods obtained in different segments to estimate the confidence on the resulting period.
The number of segments in which the period was found with a confidence level higher than 99 per cent and/or the number of segments in which the phased light-curve shows average residuals lower than its amplitude was used to estimate the robustness of the period determination.
Their approach has allowed to establish with very high confidence the rotation period for $\simeq$60 per cent of the entire catalogue (290 stars in total).

It is therefore expected that segmenting the whole Gaia time-series would be more effective for young, rapidly rotating stars, in which the ARGD time-scales are in general significantly longer than the rotation period. 

\begin{figure}
\includegraphics[angle=-90,width=85mm]{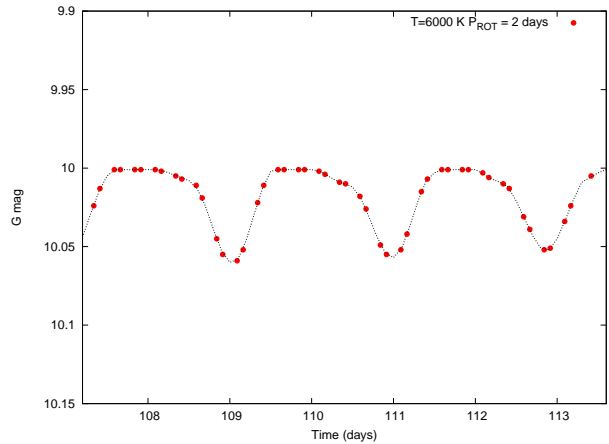}
\caption{A simulated light curve for a solar-like star with a rotation period $P~=2~d$ (solid line); points corresponding to the longest observational epoch at ($\lambda~=45^{\circ}, \beta=45^{\circ}$) are superimposed on it (red bullets). }
\label{curva}
\end{figure}

In Fig. \ref{curva} we plot a simulated light curve of a solar-like star with a rotation period $P~=2$~d  and  superimposed on it the points as sampled by the Gaia scanning law at ($\lambda=45^{\circ}, \beta=45^{\circ}$).
The sequence of 49 points  refers to the same sampling of the observations shown in Fig. \ref{ingrandimento}.
The comparison of the two figures shows how the Gaia scanning law favours  (at least at low ecliptic latitudes) the detection of the rotational modulation in fast rotating stars.
Note that in Fig. \ref{ingrandimento} the sequence of 49 points covers only one fourth of the Sun rotation period whereas in Fig. \ref{curva} it  traces three consecutive stellar rotations without the daily gaps typical of ground-based observations.

It is therefore evident that running the period search algorithms on a single group of observations like that shown in Fig. \ref{curva} rather than on the whole time-series would be more effective because the signal is more coherent along such a short-term sequence than in the whole 5-yr time-series.

\begin{figure}
\includegraphics[width=85mm]{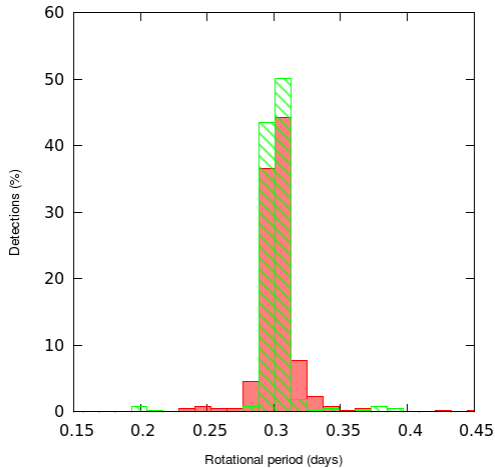}
\caption{Red histogram: distribution of periods detected by running the Lomb-Scargle algorithm on the whole time-series. Green histograms: period distribuiton obtained by running the Lomb-Scargle algorithm on sub-series. Both distributions are centered on the correct period but the distribution in green is sharper, indicating a more accurate measure of the rotation period. }
\label{confronto}
\end{figure}

We therefore performed a test to verify if, for rapidly rotating stars, the rate of correct detections can be increased by processing such short sub-series.
We selected simulated time-series for stars with rotation periods $P~=0.3$~d and placed them at ecliptic coordinates where there is at least one group of observations with $N~\ge~10$: in such a way we obtained a sample of 377 time-series.
Then, we extracted the  groups of observations satisfying the condition $N~\ge~10$ from each time-series and  ran the Lomb-Scargle algorithm on such sub-series.
In Fig. \ref{confronto} we plot the distribution of the periods detected by running the Lomb-Scargle algorithm on the whole time-series (histogram in red) as well as the distribution obtained by running the algorithm on the sub-series (in green).
Both distributions are centred on the correct period but the distribution plotted in green is sharper than the other one. 
Therefore, running the Lomb-Scargle algorithm on the short sub-series gives a more accurate period estimation.

It is interesting to notice that about 200 of the 377 analysed time-series have two or more groups of observations satisfying the requirement $N~\ge~10$. 
For fast rotators placed at these sky positions, therefore, 2-5 measures of rotation period separated by at least 6 months in time are available.
If these stars have a solar-like magnetic cycle and the latitudes of their ARs change with time, the measure of the rotation period at different epochs could even allow us to detect a possible differential rotation (cf. \citealt{ba85}).

\begin{figure}
\includegraphics[angle=-90,width=85mm]{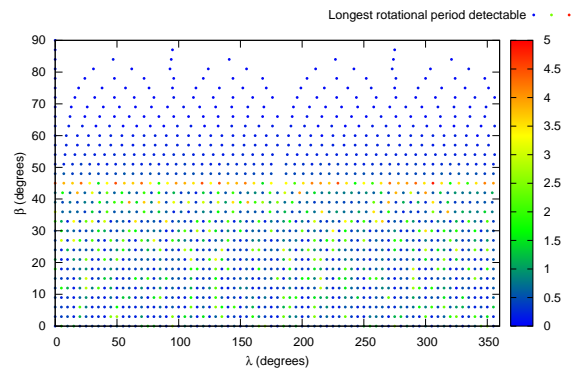}
\caption{The longest detectable rotation period versus the simulated sky directions.}
\label{upperperiod}
\end{figure}

The longest rotation period that we can detect with confidence in the Gaia case along a given sky direction by processing short-term sub-series can be roughly estimated as follows.
If we assume that a period is detectable if at least one and half consecutive rotations are sampled, then, along a given direction ($\lambda_0, \beta_0$),  the upper limit for a detectable period is $P_{\max}~\sim~L~/~1.5$~d where $L$ is the duration in days of the longest group of observations at ($\lambda_0, ~ \beta_0$). 
In Fig. \ref{upperperiod} we plot $P_{\max}$ as a function of ecliptic coordinates.
The figure shows that very fast rotators ($P~\le~1$~d) can be characterised in about 25 per cent of the simulated sky directions.
At ecliptic latitudes close to $\beta~=~45\degr$, rotation periods up to $\sim~5$~d can be recovered.

\subsubsection{Dependence on the ecliptic latitude}

\begin{figure}
\includegraphics[angle=-90,width=85mm]{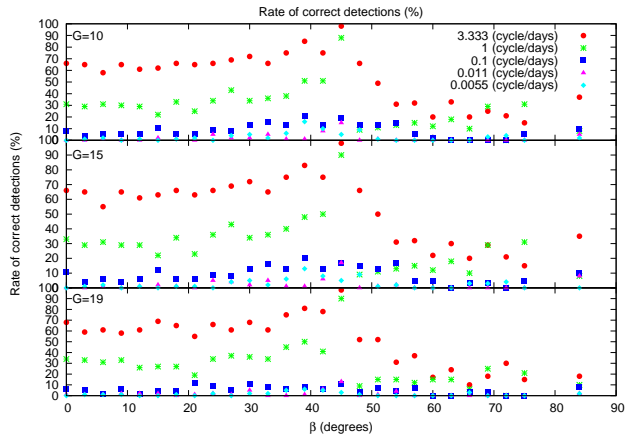}
\caption{Rate of correct period detections as a function of the ecliptic latitude $\beta$  and  the simulated frequency.}
\label{eyer}
\end{figure}

In Fig. \ref{eyer}, we plotted the rate of correct detection as a function of the ecliptic latitude for some of the simulated periods.
The periods used  correspond to the frequencies used in  Fig. 10 of \citet{EMI}, thus a comparison between our  work anf theirs can be done.

In \citet{EMI}, the rate of correct detection of strictly periodic signals is very high for all the simulated frequencies and almost independent of the ecliptic latitude: for a signal-to-noise ratio $S/N~=1.3$ it is  close to 100 per cent for all the frequencies and for $S/N=0.9$ it is greater than 40 per cent.
On the other hand, in the case of solar-like variability, the rate of correct detection is always lower than 60 per cent and is very sensitive to the ecliptic latitude.
For example, Fig. \ref{eyer} shows that for stars with a rotation period $P~=0.3$~d, the rate  increases with ecliptic  latitude in the range $0\degr~\le~\beta~\le~45\degr$; it is about 40 per cent at low ecliptic latitudes and  reaches about 100 per cent at $\beta~= 45\degr$.
Finally, it abruptly falls to about 20 per cent at ecliptic latitudes $\beta~\ge~50\degr$. 
The detection rate is also very sensitive  to the simulated periods and for slow rotators ($P~\ge~10$~d) it is always lower than 20 per cent.
The results of \cite{EMI} do not apply to solar-like variability because the amplitude and the phase of the flux modulation change with time with the  evolution of the active regions and the level of magnetic activity.
Such a behaviour is clearly shown  by the simulated light curves plotted in Fig. \ref{outputsimul} where the depth of the minima changes with time.
These light curves are more complex than a simple sinusoidal signal and two points having the same phase may have  different amplitudes if their separation in time is large enough.
Ideally, the detection of the rotation period for these stars would require a group of observations distributed  as  in Fig. \ref{curva}.
Such groups,  characterised by at least 10 observations distributed in one or a few days, are concentrated at $-45^{\circ} \le \beta~\le~45^{\circ}$ (cfr. Fig. \ref{npoints}),   and this explains why the rate of correct detections, for fast rotators, increases in this range of latitudes.

\section{The contribution of Gaia to the study of  stellar rotation in open clusters}

In the last years, several projects have been carried out to monitor late-type stars in open clusters of different ages  with the aim of measuring their rotation periods and  investigating the dependence of the stellar rotation on age, spectral type and magnetic activity.
Such projects, e.g. the RACE-OC project (Rotation and Activity Evolution in Open Cluster) \citep{RACE} or the Monitor project \citep{monitor}, are important to constrain theoretical models of  the rotational evolution of stars.
Although these studies improved our knowledge of the rotational evolution, ground-based monitoring of open clusters suffers from several limitations.
The first  is the uncertainty of the  stellar ages. 
The primary method  to estimate the age of an open cluster relies on comparing theoretical isochrones with the colour-magnitude diagram (CMD) of the cluster.
This method is very sensitive to the uncertainty in the cluster distance and, moreover, it  is affected by the contamination of the CMD by field stars.
Other limitations are related to the measure of the rotation periods, with a bias towards slow rotators ($P~\ge~10$~d), and the aliasing due to the day-night duty cycle of ground-based observations.

Gaia will allow us to improve the estimate of  open cluster ages mainly because it will supply very accurate measurements of the stellar distances.
Parallaxes measured by Gaia will have an uncertainty of only $0.1-0.3$ per cent for about
$10^5$ of the observed sources and an uncertainty $\le~1-3$ per cent for about $10^7$ sources. 
This implies, for instance, that at the end of the Gaia mission {\it individual} parallaxes will be determined with an uncertainty $\le~0.1-0.2$ per cent for all stars belonging to the Pleiades, while its {\it average} parallax for the Pleiades is presently known with an uncertainty of about 8 per cent. 
The availability of parallaxes and proper motions (and also of radial velocities for stars down to $V~\sim~15$) will allow us to assess the membership and  obtain uncontaminated CMDs for all the observed open clusters, so the expected uncertainty in their absolute ages  will become $\le~5$ per cent.

\begin{figure}
\includegraphics[angle=-90,width=85mm]{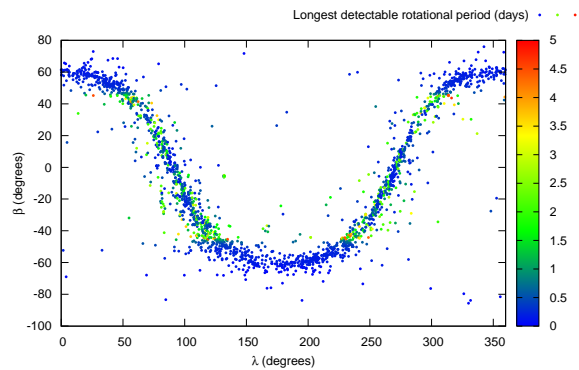}
\caption{The longest detectable rotation period is plotted for each of the open cluster listed in \citet{dias}.}
\label{mappaopen}
\end{figure}

Together with distances and ages, Gaia will supply accurate measurements of the rotation periods for fast rotating stars belonging to the open clusters in the range $-45\degr~\le~\beta~\le~45\degr$ (i.e. for about  50 per cent of the known clusters), plus less accurate measurement for longer periods.
In Fig. \ref{mappaopen},  we plotted the ecliptic coordinates of the open clusters as listed in the catalogue "Optically visible open clusters and Candidates" \citep{dias} using a colour scale to depict the longest rotation period detectable $P_{\max}$  as a function of the cluster coordinates.
Note that  most  of the clusters listed in the catalogue have angular dimensions smaller than those of the Gaia astrometric fields of view and, therefore, the pattern of the  scanning law  will not change significantly for the stars belonging to the same cluster.
Although the information on rotation periods in open clusters will not be complete, since accurate periods will be available with confidence only for the faster rotators, the number of period measurements will be unprecedented.
One can therefore envisage supplementing such an information with ground-based monitoring programs, like the ones mentioned above, to build up complete samples of rotation periods for a large number of open clusters.

\section{Conclusions}

In the present paper we study the capability of detecting the rotation periods of solar-like stars through the analysis of simulated photometric time-series that mimic those of the upcoming Gaia mission.

We test different period search methods and show that the Lomb-Scargle algorithm is the most efficient.

We show that the \cite{fap} false alarm probability (FAP) analytical formulae do not reproduce accurately the empirical CDFs constructed by means of Monte Carlo simulation unless an {\it ad hoc} evaluation of the number of independent frequencies $M$ is made.
We obtain a fit to the empirical CDFs using the \cite{fap} formula using $M$ as the fitting parameter with an accuracy better than 5 per cent.

We show that the rate of correct period detections strongly depends on the ecliptic latitude of the observed source because the time sampling generated by the Gaia scanning law exhibits different patterns at different ecliptic latitudes.
A high percentage of detections are expected in the latitude range  $-45\degr~\le~\beta~\le~45\degr$. 
The highest rate is expected at  $\beta~=~\pm~45\degr$ where the Gaia scanning law gives the highest number of transits.

The detection rate depends also on the stellar rotation period and, in particular, the Gaia scanning law is mostly suited to detect periods of rapidly rotating stars ($P~\le~5$~d). 
On the other hand, the rate of correct detections for solar-twins is, on average, $\sim~5$ per cent with a maximum of $15 - 20$ per cent at ecliptic latitudes close to $\pm~45\degr$.

The rate of correct detections is lower than the expected rate for strictly periodic variables as determined by \cite{EMI} because the light-curves of solar-like stars are more complex than a simple sinusoidal signal and the flux variations due to the rotational modulation can be hidden by the intrinsic evolutions of ARs.

The simulations described in this paper can also be used to estimate the rate of correct period detection in other surveys and, therefore, it could be a very useful tool to estimate the completeness of rotation period distributions retrieved by them.
\citet{irwin}, for example, estimated the completeness of the rotation period distributions retrieved for the open cluster M34 by running  the detection algorithms on simulated sinusoidal light curves.
The comparison between our analysis and that performed by \citet{EMI} shows that the use of a sinusoidal signal to simulate solar-like variability could significantly overestimate the rate of correct detections and, therefore, the completeness of rotation period distribution.

\section*{Acknowledgments}
This work has been partially funded by ASI under contract to INAF I/058/10/0 (Gaia Mission - The Italian Participation to DPAC).
AJK aknowledges support by the Swedish National Space Board, KE support by the Swedish Reaserch Council.

\label{lastpage}

\end{document}